\documentstyle[12pt,axodraw]{article}

\newcommand{\geqsim}{\,\raisebox{-0.6ex}{$\buildrel > \over \sim$}\,}

\def\hc{{\rm h.c.}}
\def\ibid{{\it ibid}\,}

\def\P{{\rm Pl}}

\def\gev{\,{\rm GeV}}

\def\dx{{\rm d}^4 x}
\newcommand{\dk}[1]{ \frac{{\rm d}^4 k_{#1}}{(2\pi)^4} }

\def\dt{{\rm d}^4\theta}

\def\ord{{\cal O}}
\def\be{\begin{equation}}
\def\ee{\end{equation}}
\def\ba{\begin{eqnarray}}
\def\ea{\end{eqnarray}}
\begin{document}

\thispagestyle{empty}
\begin{flushright}
{\tt ULB-TH-96/16\\ hep-ph/9609323}
\end{flushright}
\vspace{5mm}
\begin{center}
  {\Large \bf Destabilising Divergences in the NMSSM }\\
\vspace{15mm} {\large S.~A.~Abel}\\
\vspace{1cm}
{\small\it Service de Physique Th\'eorique, 
Universit\'e Libre de Bruxelles\\
Boulevard du Triomphe, Bruxelles 1050, Belgium }
\end{center} 
\vspace{2cm}

\begin{abstract}
\noindent The problem of destabilising divergences is discussed 
for singlet extensions of the MSSM. It is shown that models which 
possess either gauged-$R$ symmetry or target
space duality at the Planck scale are able to circumvent this problem
whilst avoiding cosmological domain walls. 
\end{abstract}

\pagestyle{plain}
\newpage
\section{Introduction}

There has lately been some interest in the problem of how to
accommodate an extra gauge singlet field into the minimal
supersymmetry standard model (MSSM). This is the simplest
extension which is consistent with a lightest higgs boson whose mass
exceeds the upper bound found in the MSSM~\cite{mssm}. Previously it
was thought that, by acquiring a
vacuum expectation value of $\ord (M_W)$, such a singlet could also
provide a simple solution to a fine-tuning problem in the MSSM,
the so-called `$\mu$--problem'~\cite{muprob,gm}. Because of
difficulties with cosmology (specifically the appearance of domain
walls) this now no longer appears to be the case~\cite{aw,us}. In
fact, it was shown in ref.\cite{us} that models with singlets are likely to
require symmetries in addition to those in the MSSM if they are to
avoid problems with either domain walls or fine-tuning. In this
respect models with gauge singlets are singularly {\em less} efficient
at solving fine-tuning problems. However since they allow for more
complicated higgs phenomenology, it is still worth pursuing them.
This paper concentrates on the task of building an MSSM extended by
a singlet, which avoids reintroducing the hierarchy problem,
fine-tuning, {\em and} domain walls. 

Let us take as our starting point a low-energy effective theory
which includes all the fields of the MSSM, plus one additional
singlet $N$. The superpotential is assumed to be the standard MSSM
Yukawa couplings plus the higgs interaction
\be
\label{superpot}
W_{\rm higgs}=\mu H_1 H_2 + \mu' N^2 +
\lambda{N}H_{1}H_{2}-\frac{k}{3}N^3,
\ee
and the soft supersymmetry breaking terms are taken to be
of the form  
\begin{eqnarray}
 V_{\rm soft higgs} &=
  &B \mu h_1 h_2 + B' \mu' n^2 +
\lambda A_{\lambda}nh_1h_2- 
\frac{k}{3} A_k n^3 + \hc \nonumber\\
  &&+ m^2_1 |h_1|^2
  + m^2_2 |h_2|^2
  + m^2_N |n|^2,
\end{eqnarray} 
where throughout scalar components will be denoted by lower case
letters.  For the moment let us put aside the question of how the
$\mu$ and $\mu'$ terms get to be so small (i.e. $\ord(M_W)$ instead of
$\ord(M_\P)$), and return to it later. From a low-energy point of view
the only requirement is that the additional singlet should
significantly alter the higgs mass spectrum.  This means that
$\lambda\neq 0$. There are four possibilities which can arise:

If all the other operators are absent, then in the low energy
phenomenology there is an apparent (anomalous) global ${\tilde{U}}(1)$
symmetry (orthogonal to the hypercharge), which leads to a
massless goldstone boson. Generally one expects significant 
complication to be required in
order that axion bounds are satisfied.

There are two cases which lead to a discrete symmetry.  These are $\mu
=0$, $k=0$ which leads to a $Z_2$ symmetry, and $\mu =0$, $\mu' =0$
which leads to a $Z_3$ symmetry. The latter is usually 
referred to as the next-to-minimal supersymmetric standard model 
(NMSSM)~\cite{nmssm,ellis}, and has been the main focus of work on 
singlet extensions of the MSSM.
Thus the second possibility is that there is an {\em exact} discrete
symmetry, and thus a domain wall problem associated with the existence
of degenerate vacua after the electroweak phase transition. Weak scale
walls cause severe cosmological problems (for example their density
falls as $T^2$ whereas that of radiation falls as $T^4$ so they
eventually dominate and cause power law inflation)~\cite{us}. This is
not true however, if the discrete symmetry is embedded in a broken
gauge symmetry. In this case the degenerate vacua are connected by a
gauge transformation in the full theory~\cite{ls}. After the
electroweak phase transition, one expects a network of domain walls
bounded by cosmic strings to form and then collapse \cite{ls}.
As discussed in ref.\cite{meme} bounds 
from primordial
nucleosynthesis (essentially on the reheat temperature after
inflation) require that the potential be very flat.
In addition this mechanism depends rather strongly on
the cosmology, and so models with discrete symmetry (such as
the NMSSM) remain questionable.

The third possibility is that the discrete symmetry is
broken~\cite{zko} by gravitationally suppressed
interactions~\cite{ellis,rai}.  This was the case considered and
rejected in ref.\cite{us}.  Here the very slight non-degeneracy in the
vacua, causes the true vacuum to dominate once the typical curvature
scale of the domain wall structure becomes large enough. However one
must ensure that the domain walls disappear before the onset of
nucleosynthesis and this means that the gravitationally suppressed
terms must be of order five. It was shown in ref.\cite{us} that, no
matter how complicated the full theory (i.e. including gravity), there
is {\em no} symmetry which can allow one of these terms, whilst
forbidding the operator $\nu N$, where $\nu$ is an effective
coupling. Furthermore, any such operator large enough to make the
domain walls disappear before nucleosynthesis generates these terms at
one loop anyway (with magnitude $\sim M_W^2 M_\P N$), even if they are
set to zero initially. This constitutes a reintroduction of the
hierarchy problem as emphasised in ref.\cite{destab} and as will be 
clarified in the following section.

The final case which is the subject of this paper, is when there is no
discrete symmetry at the weak scale (exact or apparent). This is true
when either $\mu\neq 0$ or both $\mu'\neq 0$ and $k\neq 0$. 
It is well known that (as in the previous case) this type of model can 
lead to dangerous divergences 
due to the existence of tadpole diagrams. Such divergences have the potential to 
destroy the gauge hierarchy unless they are either fine-tuned away, or 
removed by some higher symmetry.
In the next section the problem is quantified for the model in eq.(\ref{superpot}),
and the dangerous diagrams identified. 
It is also shown that normal gauge symmetries are not able to forbid 
these diagrams, and that they are therefore not a good candidate for the 
higher symmetry in question. Then in sections 2 and 3, it is shown that 
models which possess gauged-$R$ symmetry and target space duality respectively,
can avoid such problems. (For the reasons discussed in
ref.\cite{herbi}, gauged $R$-symmetry~\cite{herbi,gaugedr} might be
favoured over global, although the arguments presented will apply to
either case.) 

\section{The Dangerous Diagrams}

In order to demonstrate which are the dangerous diagrams associated with the 
model of eq.(\ref{superpot}), it is convenient to use the formalism of $N=1$ 
supergravity~\cite{sriv}. In this section the formalism will be described, 
and some specific examples given. Using standard power counting rules, some general
observations will then be made about the divergent diagrams. 

For completeness, let us first summarize the pertubation theory
calculation of the offending, divergent diagrams~\cite{sriv,destab}. 
The lagrangian of $N=1$ supergravity depends only on the K\"ahler function,
\be
{\cal G} = K(z^i,z^{\overline{i}}) + \ln |\hat{W}(z^i)|^2
\ee
where $z^i$ is used to denote a generic chiral superfield (visible or
hidden), and $z^{\overline{i}}=\overline{z}^i$.  Although the holomorphic 
function $\hat{W}$ is referred to
as the superpotential, it does not necessarily correspond to the
superpotential in the low energy (i.e. softly-broken, global
superymmetry) approximation. This point will be important later; hence
the hat on this superpotential. The function $K = K^\dagger $ is the
K\"ahler potential.  When supersymmetry is spontaneously broken,
divergent diagrams are most efficiently calculated using the augmented
perturbation theory rules described in ref.\cite{destab}
which are as follows.
The breaking of supersymmetry is embodied in $\theta $ and $\overline{\theta}$ dependent, 
classical VEVs for the chiral compensator, $\phi $, and K\"ahler
potential which take the form 
\ba
\label{obvious}
\phi &\sim & 1 + \frac{M_S^2}{M_\P} \theta^2 \nonumber\\ 
e^{-K/3 M^2_\P} &\sim & 1 + \frac{M_S^2}{M_\P} \theta^2 +
\frac{M_S^2}{M_\P} \overline{\theta}^2
+ \frac{M_S^4}{M_\P^2} \theta^2 \overline{\theta}^2,
\ea
where $M_S$ is the scale of supersymmetry breaking in the hidden sector, of 
order $M_S^2 \sim M_W M_\P$. (The precise forms, which are not important here, may 
be found in ref.\cite{destab}.)
Generally, in addition to renormalisable terms, the  K\"ahler potential and superpotential 
are expected to contain an infinite number of non-renormalisable terms
suppressed by powers of $M_\P$. 
There are therefore  two types of vertex which can appear in diagrams; those coming from 
the dimension-3, $\hat{W}$ operators of the form 
\be
\phi^3 \hat{W}_{ij...},
\ee
and those coming from dimension-2, $K$ operators, of the form 
\be
\phi \overline{\phi} \left( -3 e^{-K/3 M^2_\P}\right)_{ij\overline{k}\overline{l}...},
\ee
for a vertex with $z^i,z^j,z^{\overline{k}},z^{\overline{l}}...$
exiting. Here the indices $ij\overline{k}\overline{l}...$ denote covariant 
differentiation (with respect to K\"ahler transformations), so that
\ba
D_i \hat{W} & =& e^{-K/M_\P^2} \partial _i e^{K/M_\P^2}  \nonumber\\
\hat{W}_{ij}& =&D_j \hat{W}_i - \Gamma^k_{ij} \hat{W}_k 
\ea
 where $\Gamma^k_{ij}$ is the connection of the K\"ahler manifold 
described by the metric $\partial_i \partial_{\overline{j}}K$.
In order to calculate the divergent diagrams, one may now use
global superspace perturbation rules. In particular,
using the standard definitions for $D_\alpha$ and
$\overline{D}^{\dot{\alpha}}$ operators~\cite{sriv}, 
a $K$-vertex with $m$ chiral legs and $n$ antichiral legs throws 
$m$ of the $-\overline{D}^2 /4 $ and $n$ of the $-D^2 /4 $ operators onto the 
surrounding propagators. On the other hand  
a chiral vertex with $n$ chiral legs throws only $n-1$ of the $-\overline{D}^2 /4 $ operators 
onto the surrounding propagators and similarly for antichiral with $-D^2 /4 $ operators
(the difference being due to the conversion of integrations to full
superspace ones).  
The propagators are as follows~\cite{destab},
\ba
\langle z^i z^{\overline{j}} \rangle & = & 
K^{i\overline{j}} P_1 \frac{e^{K(\theta, \overline{\theta '})/3}}{\phi(\theta) \overline{\phi}
(\overline{\theta '})}
\frac{\delta^4 (x-x') \delta^4 (\theta -\theta ')}{\Box } \nonumber\\
\langle  z^{\overline{i}} z^j \rangle & = & 
K^{\overline{i}j} P_2 \frac{e^{K(\theta ',
\overline{\theta })/3}}{\phi(\theta ') \overline{\phi} (\overline{\theta })}
\frac{\delta^4 (x-x') \delta^4 (\theta -\theta ')}{\Box },
\ea
where $P_1$ and $P_2$ are the chiral and anti-chiral projection operators 
\ba
P_1 & = & \frac{ D^2 \overline{D}^2 }{16\Box} \nonumber\\
P_2 & = & \frac{ \overline{D}^2 D^2 }{16\Box} ,
\ea
and where 
\be
 \delta^4 (\theta -\theta ')=(\theta -\theta ')^2 (\overline{\theta} -\overline{\theta '})^2  .
\ee 
Since we are only interested in determining the leading divergences,
it is quite sufficient to use the massless approximation here. 

This completes our review of the perturbation theory rules. Now let us
consider the NMSSM, in which the renormalisable part of k\"ahler
potential has the canonical form, 
\be
K= z^i z^{\overline{j}}\delta_{i\overline{j}} + K_{\rm non-renorm} 
\ee
and the superpotential is of the following form; 
\be
\label{superpot2}
\hat{W}_{\rm higgs}=
\lambda{N}H_{1}H_{2}-\frac{k}{3}N^3 + \hat{W}_{\rm non-renorm} . 
\ee
The extra terms, which represent possible higher order, non-renormalisable
operators, are the terms which we are going to examine. As a
warm-up exercise, consider the case where there are no
non-renormalisable operators in $K$, and only a single non-renormalisable
coupling in the superpotential of the form 
\be
\label{superpot3}
\hat{W}_{\rm non-renorm}=
\frac{\lambda'}{M_\P} (H_{1}H_{2})^2 . 
\ee
One may hope that by adding such a coupling it is possible to remove
the domain walls which would otherwise form due to the global 
$Z_3$ symmetry apparent in the renormalisable part of
eq.(\ref{superpot2}). However, as discussed
in ref.\cite{us}, {\em there is no sufficiently large,
non-renormalisable operator that can be added to the superpotential,
which does not destabilise the gauged hierarchy}. Here `sufficiently
large' means that the cosmological walls must disappear before the onset of
primordial nucleosynthesis for which one requires $\lambda'\geqsim 10^{-7}$. 
For the operator in question, this is due to the 3-loop diagram in
fig.(1), which gives rise to a contribution to the effective action of the form,
\ba
\label{3-loop}
\delta S &=& \frac{-k\lambda' \lambda^2}{M_\P}
\int \dx_1 \ldots \dx_4 \dt_1\ldots \dt_4
N(x_1,\theta_1) \frac{\phi(\theta_1)}{\phi(\theta_4)}  
{ e^{K_{(12)}/3}e^{K_{(13)}/3}e^{2 K_{(42)}/3}e^{2 K_{(43)}/3}}
\nonumber\\
& & \hspace{1cm}\times
\left( \frac{\overline{D}^2_1 \delta_{12}}{4\Box_1}\right)
\left( \frac{          D ^2_2 \delta_{24}}{4\Box_2}\right)
\left( \frac{\overline{D}^2_4 \delta_{43}}{4\Box_4}\right)
\left( \frac{          D ^2_3 \delta_{31}}{4\Box_3}\right)
\left( \frac{D_2^2 \overline{D}^2_2 \delta_{24}}{16\Box_2}\right)
\left( \frac{\overline{D}^2_4 D^2_4 \delta_{43}}{16\Box_4}\right),
\ea
where $\delta_{ij}=\delta^4 (x_i-x_j) \delta^4 (\theta_i -\theta_j) $,
and here $K_{(ij)}=K(\theta_i,\overline{\theta}_j )$.\\
\begin{picture}(375,250)(0,40)
\Line(200,250)(238,167)
\Line(200,250)(162,167)
\Line(200,150)(200,100)
\Vertex(200,150){3}
\Vertex(200,250){3}
\Vertex(238,167){3}
\Vertex(162,167){3}
\CArc(200,200)(50,270,90)
\CArc(200,200)(50,90,270)
\Text(200,160)[]{\scriptsize $1$}
\Text(200,260)[]{\scriptsize $4$}
\Text(238,180)[]{\scriptsize $3$}
\Text(162,180)[]{\scriptsize $2$}
\Text(195,100)[]{\scriptsize $N$}
\Text(142,208)[]{\scriptsize $H_2$}
\Text(172,208)[]{\scriptsize $H_1$}
\Text(212,208)[]{\scriptsize $H_2$}
\Text(242,208)[]{\scriptsize $H_1$}
\Text(170,153)[]{\scriptsize $N$}
\Text(230,153)[]{\scriptsize $N$}
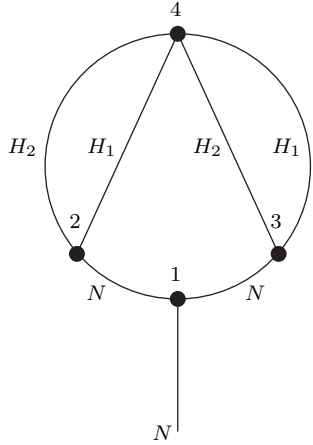
\Text(200,75)[]{ figure 1: Divergent tadpole diagram from $(H_1 H_2)^2$ operator.}
\end{picture}\\
\noindent
One can evaluate this expression by integrating by parts to expose 
factors of $\delta^4 (\theta_i-\theta_j)$ and thus eliminating 
$\theta$ integrals in the standard manner. Acting on the $\phi $ or
$e^{K/3}$ factors always reduces the degree of divergence as is
obvious from eqn.(\ref{obvious}). Factors of $D^2 \overline{D}^2 $ 
may be removed using the identities,
\ba
\label{didents}
D^2 \overline{D}^2 D^2 &=& 16 \Box D^2 \nonumber \\
\overline{D}^2 D^2 \overline{D}^2 &=& 16 \Box \overline{D}^2 \nonumber \\
16 &=&\int \dt_2 \delta^4(\theta_2-\theta_1) D^2 \overline{D}^2
\delta^4(\theta_2-\theta_1) \nonumber\\
16 &=&\int \dt_2 \delta^4(\theta_2-\theta_1) D^2 \overline{D}^2
\delta^4(\theta_2-\theta_1).  
\ea
The integral is reduced to a single 
integral over $\theta_1$ of the form,
\be
\delta S = \frac{-2 k\lambda' \lambda^2}{M_\P}\int \dx_1 \ldots \dx_4 \dt_1
N(x_1,\theta_1)  
 e^{2 K_{(11)}} 
\left( \frac{\delta^4 x_{31}}{\Box_3}\right)
\left( \frac{\delta^4 x_{43}}{\Box_4}\right)^2
\left( \frac{\delta^4 x_{24}}{\Box_2}\right)^2
\delta^4 x_{12} ,
\ee
where $\delta^4 x_{ij}=\delta^4 (x_i-x_j) $.
Converting the delta functions to momentum space, one finds a
contribution to the effective action of 
\be
\delta S = -2 k\lambda' \lambda^2\int \dx_1 \dt_1
N(x_1,\theta_1)  
 e^{2 K_{(11)}} I_3,
\ee
in which $I_3$ is the quadratically divergent 3-loop integral,
\be 
I_3=\int \dk{1}\dk{2}\dk{3} \frac{1}{k_1^2 k_2^2 k_3^2 (k_1-k_2)^2
(k_1-k_3)^2 }= \ord (M_\P^2 /(16\pi^2 )^{3}) ,
\ee
where the integral has been regularised with a cut-off of order $M_P$. 
Inserting the $\theta$ dependent VEVs of eqn.(\ref{obvious}) into the
above, results in terms in the effective potential of the form 
\be
\delta V\approx \frac{2 k\lambda' \lambda^2}{(16 \pi^2)^3}
\left( (n+n^*) M_\P M_W^2 + (F_N + F_N^*) M_\P M_W\right)  
\ee
which clearly destabilises the hierarchy unless $\lambda'$ is
sufficiently small, so small in fact that it is unable to remove the 
cosmological domain walls before the onset of nucleosynthesis~\cite{us}.
The non-renormalisable term in eq.(\ref{superpot3}), is
(to leading order in $M_\P^{-1}$) equivalent to adding instead the term
\begin{equation} 
 K_{\rm non-renorm} = 
- \frac{\lambda'}{\lambda} \left(\frac{N^{\dagger}H_{1}H_{2} +
\hc}{M_{\P}}\right)
- \frac{k \lambda'}{\lambda^2} \left(\frac{N^{\dagger}H_{1}H_{1}^\dagger +
\hc}{M_{\P}}\right) ,
\end{equation}
in the K\"ahler potential. This may be seen by making the redefinitions
\ba
 N &\rightarrow  & N - \frac{\lambda' H_1 H_2}{\lambda M_\P} \nonumber\\
 H_1 &\rightarrow  & H_1 - \frac{\lambda' k N H_1}{\lambda^2 M_\P}.
\ea
This provides a useful check of the perturbation theory rules. The
divergent diagrams in the redefined model are of the form shown in
fig.(2), where black vertices are chiral and white ones come from 
the $K_{\rm non-renorm}$ terms in the K\"ahler potential. \\
\begin{picture}(450,125)(-50,30)
\CArc(-25,100)(20,270,90)
\CArc(-25,100)(20,90,270)
\Line(-25,80)(-25,70)
\BCirc(-25,80){2}
\Text(12,100)[]{+}
\CArc(50,100)(20,270,90)
\CArc(50,100)(20,90,270)
\Vertex(30,100){2}
\Vertex(50,80){2}
\Line(30,100)(70,100)
\Line(50,80)(50,70)
\BCirc(70,100){2}
\CArc(125,100)(20,270,90)
\CArc(125,100)(20,90,270)
\Vertex(105,100){2}
\Vertex(145,100){2}
\Line(105,100)(145,100)
\Line(125,80)(125,70)
\BCirc(125,80){2}
\Text(87,100)[]{+}
\CArc(200,100)(20,270,90)
\CArc(200,100)(20,90,270)
\Vertex(183,110){2}
\Vertex(217,110){2}
\Vertex(183,90){2}
\Vertex(200,80){2}
\Line(200,80)(200,70)
\Line(183,110)(217,110)
\Line(183,90)(217,90)
\BCirc(217,90){2}
\Text(163,100)[]{+}
\CArc(275,100)(20,270,90)
\CArc(275,100)(20,90,270)
\Vertex(258,110){2}
\Vertex(292,110){2}
\Vertex(258,90){2}
\Vertex(292,90){2}
\Line(275,80)(275,70)
\Line(258,110)(292,110)
\Line(258,90)(292,90)
\BCirc(275,80){2}
\Text(237,100)[]{+}
\CArc(350,100)(20,270,90)
\CArc(350,100)(20,90,270)
\Vertex(370,100){2}
\Vertex(360,117){2}
\Vertex(360,83){2}
\Vertex(350,100){2}
\Line(350,100)(360,117)
\Line(350,100)(360,83)
\Line(330,100)(350,100)
\Line(370,100)(380,100)
\BCirc(330,100){2}
\Text(312,100)[]{+}
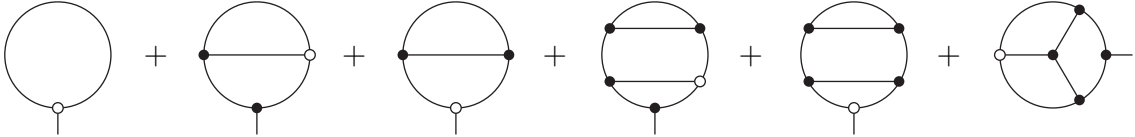
\Text(175,50)[]{figure 2: Equivalent diagrams to fig.(1) when the fields
are redefined. }
\end{picture}\\
The 1-loop divergent contributions were shown by 
Jain in ref.\cite{destab} to cancel unless the 
trilinear terms couple directly to hidden sector fields. This result
can easily be recovered here, since the diagram gives
\be
\delta S = \frac{M_\P}{2 (16 \pi^2)} \int \dx_1 \dt_1  
K_{N H_1 \overline{H}_1} K^{H_1 \overline{H}_1} N(x_1,\theta_1) + \hc 
\ee
where we have approximated 
\be
\int \dk{1} \frac{1}{k_1^2}= \ord ( M_\P^2 / (16 \pi^2) ) .
\ee
Without any direct coupling between $H_1$ and a hidden sector field, 
the VEVs of eq.(\ref{obvious}) do not appear, and the diagram does not
give dangerous terms. 
The 2-loop contributions are easily found to cancel amongst
themselves. With a little effort the remaining divergences can  
also be shown to cancel except the single (Mercedes) diagram
of fig.(3).\\
\begin{picture}(375,250)(0,40)
\Line(200,200)(238,167)
\Line(200,200)(162,167)
\Line(200,200)(200,250)
\Line(200,150)(200,100)
\Vertex(200,200){3}
\Vertex(200,150){3}
\Vertex(238,167){3}
\Vertex(162,167){3}
\CArc(200,200)(50,270,90)
\CArc(200,200)(50,90,270)
\BCirc(200,250){3}
\Text(200,160)[]{\scriptsize $1$}
\Text(200,260)[]{\scriptsize $5$}
\Text(195,205)[]{\scriptsize $4$}
\Text(238,180)[]{\scriptsize $3$}
\Text(162,180)[]{\scriptsize $2$}
\Text(195,100)[]{\scriptsize $N$}
\Text(142,208)[]{\scriptsize $H_2$}
\Text(177,190)[]{\scriptsize $H_1$}
\Text(225,190)[]{\scriptsize $H_2$}
\Text(242,208)[]{\scriptsize $H_1$}
\Text(170,153)[]{\scriptsize $N$}
\Text(230,153)[]{\scriptsize $N$}
\Text(195,225)[]{\scriptsize $N$}
\Text(190,75)[]{ figure 3 }
\end{picture}\\

The contribution of this diagram to the effective action is,
\ba
\delta S &=& \frac{-k\lambda' \lambda^2}{M_\P}
\int \dx_1 \ldots \dx_5 \dt_1\ldots \dt_5
N(x_1,\theta_1) \frac{\phi(\theta_1)}{\phi(\theta_5)}  
\nonumber\\
&& \times
e^{K_{(12)}/3}e^{K_{(13)}/3}e^{K_{(42)}/3}e^{K_{(43)}/3}
e^{K_{(45)}/3}e^{K_{(52)}/3}e^{K_{(53)}/3}
\left( \frac{\overline{D}^2_1 \delta_{12}}{4\Box_1}\right)
\left( \frac{D ^2_2 \overline{D}^2_2 \delta_{25}}{16\Box_2}\right)
\nonumber\\
& & \times
\left( \frac{\overline{D}^2_5 \delta_{53}}{4\Box_4}\right)
\left( \frac{          D ^2_3 \delta_{31}}{4\Box_3}\right)
\left( \frac{D_2^2 \overline{D}^2_2 \delta_{24}}{16\Box_2}\right)
\left( \frac{\overline{D}^2_4 D^2_4 \delta_{43}}{16\Box_4}\right)
\left( \frac{ D^2_5                 \delta_{54}}{4\Box_5}\right) .
\ea
By integrating by parts with $\overline{D}_4^2$, $\overline{D}_5^2$
and $D_5^2$, and using the rules in eqn.(\ref{didents}),
the last factor becomes simply $ \delta_{54}$. The $\langle 4 5 \rangle$
propagator effectively collapses and the integral over
$(x_5,\theta_5)$ results in eqn.(\ref{3-loop}) as required. (Again,
when evaluating the leading divergences, one
may ignore $D^2$ operators acting on $\phi $ and $e^{K/3}$.)

Having gained some confidence in calculation of divergences, we can
now go
on to systematically consider the other operators which may appear in
$\hat{W}$ or $K$. In order to determine 
exactly which ones are dangerous, let us first restrict our attention
to operators in $\hat{W}_{\rm non-renorm}$.
Obviously the degree of fine-tuning
decreases with higher order since each loop gives a factor $\Lambda^2
/(16 \pi^2 )$  where $\Lambda $ is a cut-off, and involves more Yukawa
couplings. It therefore seems reasonable to disregard contributions 
which are higher than six-loop since they 
are unable to destabilise the hierarchy. Upto and including six loop, 
the following operators are potentially dangerous if they
appear in the superpotential (multiplied by any function of 
hidden sector fields), since one can write down a tadpole diagram using
them (together with the trilinear operators of the NMSSM);
\vspace{0.5cm}
\begin{center}
\begin{tabular}{||l|l|r||}   \hline
  \mbox{Operator}     & \mbox{resp. diagram} & \mbox{Loop-order} \\ \hline\hline 
    $N^2$, $H_1 H_2$     & 3a,3a & 1         \\ \hline 
    $N^4$, $N^2 H_1 H_2$     & 3b,3b & 2         \\ \hline 
         $(H_1 H_2)^2  $, $N (H_1 H_2)^2$,
$N^3 (H_1 H_2)$, $N^5$     & 3c,3d,3d,3d &3         \\ \hline
$N^3 (H_1 H_2)^2$, $N^5 (H_1 H_2)$, $N^7$    & 3e,3e,3e,3e & 4         \\ \hline
$N (H_1 H_2)^3$, $N^2 (H_1 H_2)^3$, $N^4 (H_1 H_2)^2$,
$N^6 (H_1 H_2)$, $N^8$     & 3f,3g,3g,3g,3g & 5         \\ \hline
$N^4 (H_1 H_2)^3$, $N^6 (H_1 H_2)^2$,
$N^8 (H_1 H_2)$, $N^{10}$     & 3h,3h,3h,3h & 6         \\ \hline
\end{tabular}
\end{center}
\vspace{0.5cm}
The corresponding tadpole diagrams for each operator are shown in
fig.(4a-h). (Figure (4c) is the diagram which was evaluated above.)
Notice that, since the leading divergences involve chiral or antichiral
vertices only, an operator must break the $Z_3$ symmetry in 
$\hat{W}$ in order for it to be dangerous (so that for example 
$N^2 (H_1 H_2)^2 $ does not destabilise the hierarchy). 
The first two operators are the exception in this list, 
since one cannot say with certainty whether or not their contributions to the 
effective potential will be dangerous. This depends on how the couplings $\mu $ 
or $\mu'$ are generated. Specifically, the diagram in fig.(4a)
generates logarithmically divergent terms of the form 
\be
\delta V = \frac{\log \Lambda^2}{32 \pi^2}
 \int \mbox{d}^4\theta e^{2 K/3 M_\P^2} \varphi\overline{\varphi}
\hat{W}_{ij} \overline{\hat{W}}^{ij} +\ldots
\ee
These are the divergent terms which lead to logarithmic
running of the soft-breaking scalar masses. However, if there is a 
$\mu$-term produced directly in the superpotential from some product
of hidden sector fields ($\mu = \Phi^m/M_\P^{m-1}$ for
example), the contribution above includes
\be
 \frac{\log \Lambda^2}{32 \pi^2}
 \int \dt \mu (\Phi) \lambda^\dagger N^\dagger = 
 \frac{\log \Lambda^2}{32 \pi^2} \lambda^\dagger F_N^\dagger  
\frac{m \phi^{m-1} F_\Phi}{M^{m-1}_\P}
\sim \left( \frac{M_\P}{M_W}\right)^{1/m}M_W^{2} F_N^\dagger . 
\ee
where since $\Phi$ is a hidden sector field, one can assume that 
$F_\Phi \sim M_W M_\P $, and that also  
$\langle |\phi |^m \rangle \sim M_W M_\P^{m-1}$ in
order to get $\mu \sim M_W $. This leads to a value of $F_N \gg M_W$
unless $m$ is extremely large, destabilising the gauge hierarchy.
If $\mu $ is generated in the visible sector on the other hand, it may be 
possible to avoid this conclusion\footnote{I would like to thank G.~G.~Ross 
for pointing this out.}. In this sense such terms have the same status as
the trilinear couplings in the K\"ahler potential which were discussed above. 

It has already been demonstrated that the next three operators will 
lead to dangerous divergences and must be forbidden. Not all of the 
remaining operators are dangerous however. Consider for instance
adding a dimension-7 operator to the superpotential;
\be
\hat{W}_{\rm non-renorm} = \frac{\lambda'}{M_\P^4} N^7 . 
\ee
In this case the (Garfield) diagram of fig.(4e) looks potentially
dangerous, since it also appears to be a divergent tadpole
contribution. Its contribution to the effective action is 
\ba
\label{garfield}
\delta S &=& \frac{k^2\lambda'}{18 M^4_\P}
\int \dx_1 \dx_2 \dx_3 \dt_1\dt_2\dt_3
N(x_1,\theta_1) \frac{1}{\phi(\theta_1)^3}  
 e^{K_{(12)}}e^{K_{(13)}}
\nonumber\\
  & & \hspace{1cm}\times
\left( \frac{D^2_2\overline{D}^2_2 \delta_{21}}{16\Box_2}\right)^2
\left( \frac{\overline{D}^2_2 D^2_2 \delta_{21}}{16\Box^2_2}\right)
\left( \frac{D^2_3\overline{D}^2_3 \delta_{31}}{16\Box_3}\right)^2
\left( \frac{-\overline{D}^2_3 \delta_{31}}{4\Box_3}\right).
\ea
\newpage
\begin{picture}(400,400)(0,50)
\CArc(50,400)(20,270,90)
\CArc(50,400)(20,90,270)
\Vertex(30,400){2}
\Vertex(70,400){2}
\Line(70,400)(80,400)
\Text(0,400)[]{ (a)}
\CArc(200,400)(20,270,90)
\CArc(200,400)(20,90,270)
\Vertex(180,400){2}
\Vertex(220,400){2}
\Line(180,400)(230,400)
\Text(150,400)[]{ (b)}
\CArc(350,400)(20,270,90)
\CArc(350,400)(20,90,270)
\Vertex(330,400){2}
\Vertex(370,400){2}
\Vertex(360,417){2}
\Vertex(360,383){2}
\Line(330,400)(360,417)
\Line(330,400)(360,383)
\Line(370,400)(380,400)
\Text(300,400)[]{ (c)}
\CArc(50,300)(20,270,90)
\CArc(50,300)(20,90,270)
\Vertex(30,300){2}
\Vertex(70,300){2}
\CArc(85,300)(15,270,90)
\CArc(85,300)(15,90,270)
\Vertex(100,300){2}
\Line(100,300)(110,300)
\Line(30,300)(70,300)
\Text(0,300)[]{ (d)}
\CArc(200,300)(20,270,90)
\CArc(200,300)(20,90,270)
\Vertex(180,300){2}
\Vertex(220,300){2}
\CArc(240,300)(20,270,90)
\CArc(240,300)(20,90,270)
\Vertex(260,300){2}
\Line(220,300)(220,280)
\Line(180,300)(260,300)
\Text(150,300)[]{ (e)}
\CArc(350,300)(20,270,90)
\CArc(350,300)(20,90,270)
\CArc(345,274)(30,63,117)
\CArc(345,326)(30,243,297)
\Vertex(330,300){2}
\Vertex(370,300){2}
\Vertex(360,317){2}
\Vertex(360,283){2}
\Vertex(360,300){2}
\Line(330,300)(320,300)
\Line(330,300)(360,317)
\Line(330,300)(360,283)
\Line(370,300)(360,300)
\Text(300,300)[]{ (f)}
\CArc(50,195)(20,17,163)
\CArc(50,205)(20,197,343)
\Vertex(30,200){2}
\Vertex(70,200){2}
\Vertex(110,200){2}
\CArc(90,195)(20,17,163)
\CArc(90,205)(20,197,343)
\Line(30,200)(110,200)
\CArc(93,180)(30,137,223)
\CArc(47,180)(30,317,403)
\Vertex(70,160){2}
\Line(70,160)(70,150)
\Text(00,200)[]{ (g)}
\CArc(200,195)(20,17,163)
\CArc(200,205)(20,197,343)
\Vertex(180,200){2}
\Vertex(220,200){2}
\Vertex(260,200){2}
\CArc(240,195)(20,17,163)
\CArc(240,205)(20,197,343)
\Line(180,200)(260,200)
\CArc(243,180)(30,137,223)
\CArc(197,180)(30,317,403)
\Vertex(220,160){2}
\Line(220,160)(220,220)
\Text(150,200)[]{ (h)}
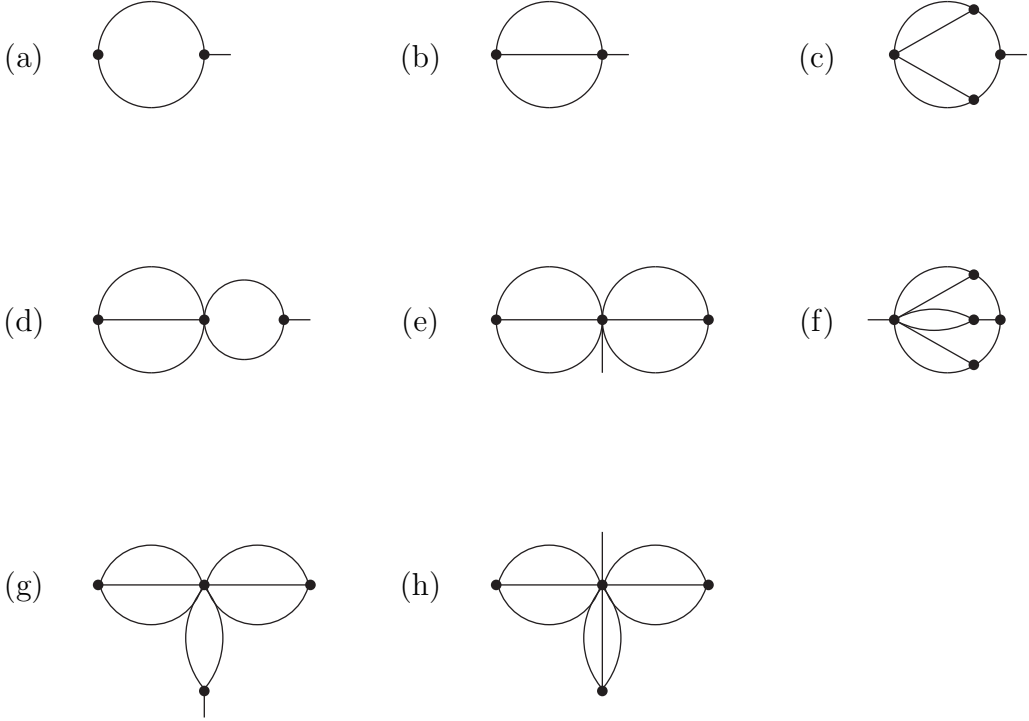
\Text(200,100)[]{figure 4: Tadpole diagrams for non-renormalisable
operators in $\hat{W}$ upto 6-loop. }
\end{picture}\\

Again by integrating by parts with $\overline{D}_2^2$ 
and $\overline{D}_3^2$ one can extract the leading term, but this 
time, one is forced to act at least once upon the $e^K$ factors, 
because in total there is an odd number of $D^2$ and $\overline{D}^2$
operators. 
The result is 
\be
\delta S = \frac{k^2\lambda'}{18 M^4_\P}\int \dx_1 \dt_1
N(x_1,\theta_1) \frac{1}{\phi(\theta_1)^3}
 \left(-\frac{\overline{D}^2}{4}e^{ 2 K_{(11)}}\right)  I_4,
\ee
in which $I_4$ is the quartically divergent 4-loop integral,
\be 
I_4=\int \dk{1}\dk{2}\dk{3}\dk{4} \frac{1}{k_1^2 k_2^2 k_3^2 k_4^2
 (k_1-k_2)^2 (k_3-k_4)^2 }= \ord (M_\P^4 /(16\pi^2 )^{4}) .
\ee
The final contribution to the effective potential is not harmful to
the gauge hierarchy;
\be
\delta V \approx \frac{-k^2\lambda'}{9 (16 \pi^2)^4}
\left( (F_N+F_N^*) M_W^2 + (n+n^*) M_W^3 \right).
\ee
This is clearly the case whenever the total number of $D^2$ and
$\overline{D}^2$ operators is odd. This fact leads one quite easily 
to the chief result of this section, which is that, for the 
model of eqn.(\ref{superpot2}), {\em any extra
odd-dimension operators in $\hat{W}$ or even-dimension operators in 
$K$ are not harmful to the gauge hierarchy.}   

This may be deduced by first generalising the supergraph, power counting 
rules. Let there be $V_d$ superpotential vertices of dimension $d+3$
(that is of the form $z^{d+3}/M_\P^d $), and $U_d$ K\"ahler potential 
vertices of dimension $d+2$ (of the form $z^{d+2}/M_\P^d $). To the 
divergence, a propagator counts as $ 1/p^2 $, a $V_d$
vertex as $p^{d+2}$ (from the $D^2$ factors on its legs), 
a $U_d$ vertex as $p^{d+2}$, and each loop variable as $p^2$. 
In addition each external chiral leg removes a $D^2$ operator of 
the vertex, effectively contributing $1/p$.   
Hence the total degree of divergence is~\cite{sriv}, 
\be
D= 2 L -2 P - E_c + \sum_d V_d (d+2) + \sum_d U_d (d+2),
\ee
where $L$ is the number of loops, $P$ is the number of propagators,
and $E_c$ is the number of external chiral legs. There are two useful
relations; the first is 
\be
\label{pident}
2 P + E_c = \sum_d V_d (d+3) + \sum_d U_d (d+2),
\ee
the right hand side being simply the number of external legs when 
there are no propagators; the second arises from counting the internal
momentum variables, one of which is removed by each vertex delta function, 
\be
P - L = \sum_d V_d + \sum_d U_d -1 .
\ee
Substituting these gives the following value for the divergence
\be
D= 2 - E_c + \sum_d V_d + \sum_d U_d.
\ee
The actual contribution to the effective potential is therefore of the
form 
\be
\delta V \sim \frac{\Lambda^{2-E_c +  \sum_d V_d + \sum_d U_d}}
{M_\P^{ \sum_d V_d + \sum_d U_d}} \sim M_\P^{2-E_c}.
\ee
This is the result of ref.\cite{sriv,destab}, which says that in 
$N=1$ supergravity, apart from a quadratic vacuum term, 
the only divergent contribution to the effective potential is linear in fields ($E_c=1$).
Now consider the total number, $N_{D^2}$, of $D^2$ and $\overline{D}^2$ operators.
There are $d+2$ from every vertex, $-1$ from every external chiral
line, and 2 on every propagator, giving
\be
N_{D^2} = 2 P - E_c + \sum_d V_d (d+2) + \sum_d U_d (d+2)
\ee
in total. In order for a diagram to be harmful, this number must be 
even, and hence when $E_c=1$, 
\be
\sum_d V_d d + \sum_d U_d d = \mbox{odd}.
\ee
This can only be satisfied if there is at least one vertex which has
an odd $d$, thus proving the statement above. 
(Substituting eq.(\ref{pident}) shows that this also means the 
total number of chiral and antichiral vertices is even.) 

The relatively restrictive constraint that the superpotential be a
holomorphic function means that there are now only 13
dangerous operators in $\hat{W}$. The K\"ahler potential is restricted
only by the condition, $K=K^\dagger$ however. Apart from the trilinear
operators (which as we
have seen above only destabilise the gauge hierarchy if they directly 
couple visible and hidden sector fields), there is a much larger number 
of higher dimension operators which must be forbidden here. For example 
the operator,
\be
K_{\rm non-renorm}=\lambda' N^{\dagger 2} N (H_1 H_2) 
\ee
leads to the diagram in Fig.(5), whose contribution to the effective
action is 
\be 
\delta S \approx -\frac{M_\P k \lambda \lambda'}{18 (16 \pi^2)^4}
\int \dx_1 \dt_1
N(x_1,\theta_1) \frac{\phi(\theta_1)}{\overline{\phi}(\overline{\theta}_1)}  
 e^{5 K_{(11)}/3},
\ee 
which again gives $n$ a VEV of $\ord (10^{11} \gev )$. Clearly {\em any} 
odd-dimension operator which breaks the $Z_3$ symmetry of
eq.(\ref{superpot3}) may appear in $K$ and will destroy the gauge
hierarchy if it does so.\\
\begin{picture}(375,250)(30,60)
\Line(250,200)(150,200)
\Line(320,200)(350,200)
\Vertex(150,200){3}
\Vertex(320,200){3}
\CArc(200,200)(50,270,90)
\CArc(200,200)(50,90,270)
\CArc(285,200)(35,270,90)
\CArc(285,200)(35,90,270)
\BCirc(250,200){3}
\Text(200,155)[]{\scriptsize $N$}
\Text(200,205)[]{\scriptsize $H_2$}
\Text(200,255)[]{\scriptsize $H_1$}
\Text(285,240)[]{\scriptsize $N$}
\Text(285,170)[]{\scriptsize $N$}
\Text(340,205)[]{\scriptsize $N$}
\Text(250,100)[]{ figure 5 }
\end{picture}\\

Hence a particularly attractive way to ensure a model with
singlets which is natural, is to devise a symmetry which forbids
odd-dimension terms in $K$, and even-dimension 
terms in $\hat{W}$. 
This is the approach taken in the next two sections. (A 
possible alternative which will not be considered here is to 
include an extra symmetry in
the visible sector, which ensures these couplings are always
suppressed by some field whose VEV is extremely small.) 

To finish this section, let us recapitulate the arguments of
ref.\cite{us} which make it clear that such a symmetry cannot 
be a normal gauge symmetry. 
For simplicity, take this to be a $U(1)_X$
symmetry (the extension to non-abelian cases is trivial), and let the 
$Z_3$ symmetry be broken by a $H_1 H_2$ or $N^2$ term in $K$. Such couplings 
provide naturally small $\mu\sim M_W$ or $\mu' \sim M_W$ in the
effective low energy global superpotential $W$~\cite{gm}. 
The other effective couplings at the weak scale are in general
arbitrary functions of hidden sector fields which carry charge under
the new $U(1)_X$ which shall be referred to collectively as $\Phi$ (with
$\xi =\Phi/M_\P$). It is simple to see that one cannot use this
symmetry to forbid terms linear in $N$.
If $\mu (\xi)\neq 0$ then $\mu (\xi)$ must have the same charge as
$\lambda (\xi) N$ and therefore $(\mu (\xi))^\dagger \lambda (\xi) N $
is uncharged. If both $\mu'\neq 0$ and $k\neq 0$ then $\mu' (\xi)$
must have the same charge as $k (\xi) N$ and therefore $(\mu'
(\xi))^\dagger k (\xi) N $ is uncharged. Once such a linear operator 
has been constructed, it is of course trivial to construct all the
other dangerous operators. 

One should bear in mind that if one sets these couplings to zero by
hand in the first place, they remain small to higher order in
perturbation theory. So this is merely a fine-tuning problem.
One might also argue that the nature of this fine-tuning problem is
different from that of the $\mu$-problem, since in the latter 
the coupling has to be very small, whereas here the couplings may 
just happen to be absent (as for example are superpotential mass 
terms in string theory).
However, the extremely large number of dangerous operators 
makes this fine tuning problem a particularly serious one.
In the next two sections, two examples are presented which are able to
avoid this problem.

\section{Models with $R$-symmetry}

The reason that it has not been possible to forbid divergent tadpole
diagrams 
in the models that have been discussed here and in ref.\cite{us}, is
that the K\"ahler potential and superpotential have the same charges
(i.e. zero).  There are however two available symmetries in which the
K\"ahler and superpotentials transform differently. These may
accommodate singlet extensions to the MSSM simply and without fine-tuning.

The first is gauged $U(1)_R$-symmetry~\cite{herbi,gaugedr}. In this case the 
K\"ahler potential has zero $R$-charge, but the superpotential has $R$-charge 
2. This means that the standard renormalisable NMSSM higgs superpotential,
\be 
\label{rwhiggs}
\hat{W}_{\rm higgs}=\lambda N H_{1}H_{2}-
\frac{k}{3}N^3,
\ee
has the correct $R$-charge if $R(N)= 2/3$ and $R(H_1)+R(H_2) =4/3 $. 
So consider the K\"ahler potential   
\be 
\label{quad}
{\cal G} = y_i y^i + \Phi \overline{\Phi}   
+ \left( \frac{\alpha}{M^2_\P}\Phi H_1 H_2 + 
 \frac{\alpha '}{M^2_\P}\Phi N^2 +\hc \right) 
+ \log |\hat{W} + g(\Phi )|^2 ,
\ee
where $y_i$ are the visible sector fields and 
where $\Phi$ represents a hidden sector field with superpotential 
$g(\Phi )$ which aquires a VEV of $\ord (M_\P)$. 
(It may represent arbitrary functions of hidden 
sector fields in what follows). This next-to-minimal choice 
of K\"ahler potential is the one proposed in ref.\cite{gm} which leads
to naturally small $\mu$ and $\mu'$ couplings in the low energy
(global supersymmetry) approximation $W$. Specifically, the terms which 
arise in the scalar potential are~\cite{gm,us}
\begin{equation}
 V_{\rm scalar} = W_{i} W^{i} + m^2 y_{i}y^{i}
                  + m \left[y^{i}W_{i}+(A-3)\tilde{W}
                  + (B-2)m \mu H_{1} H_{2} + (B-2)m \mu^\prime N^2 
+ \hc\right],
\end{equation}
where $\tilde{W}$ are the trilinear terms of the superpotential $\hat{W}$,
rescaled according to
\begin{equation}
 \tilde{W} = \langle\exp{(\Phi \overline{\Phi}/2 M_{\P}^2)}
\rangle \hat{W}.
\end{equation}
Here $W$ is the new low energy superpotential including the 
$\mu$ and $\mu'$ terms,
\begin{equation}
 W = \hat{W} + \mu H_{1} H_{2} + \mu' N^2,
\end{equation}
and $m$ is the gravitino mass
\begin{equation}
 m = \langle \exp{(\Phi \overline{\Phi} / 2 M_{\P}^2)} g^{(2)}\rangle,
\end{equation}
where $g^{(2)}$ are the quadratic terms in $g$, and 
where the VEV of $g^{(2)} = M_S^2/M_{\P}$ is set by hand such
that $M_{S}\sim 10^{11}$ GeV. 
Applying the constraint of vanishing cosmological constant, one finds 
that the universal trilinear scalar coupling, $A=\sqrt{3}
\langle \Phi/M_{\P}\rangle $, and that the bilinear couplings are given
by, 
\begin{eqnarray}
 B &= &(2 A-3)/(A-3) \nonumber\\
 |\mu| &= &\left|\frac{m\alpha(A-3)}{\sqrt{3}}\right| \nonumber\\
 |\mu^\prime| &= &\left|\frac{m\alpha^\prime(A-3)}{\sqrt{3}}\right|.
\end{eqnarray}                                                 
All dimensionful parameters at low energy are of order $M_W$.

Invariance of the K\"ahler potential requires that $R(\Phi) =- 4/3 $. 
It is easy to see that with this set of $R$-charges there
can never be odd-dimension operators in $K$, or even-dimension ones in
$\hat{W}$. Indeed the operators which can appear in the superpotential 
can be written as, 
\be
\hat{O}_{c}=
\frac{\Phi^c}{M_\P^c} \frac{y^{(d+3)}}{M_\P^d}, 
\ee
where $y$ stands for any of the visible sector fields. In order to
have $R$-charge 2, they must satisfy 
\be
\frac{2 (d+3)}{3} - \frac{4 c}{3}=2
\ee
or $d=2c$. Hence only odd-dimension operators are allowed in
$\hat{W}$. The operators which can appear in the K\"ahler potential
are of the form 
\be
\hat{O}_{abc}=
\frac{(\Phi \overline{\Phi})^b}{M_\P^{2b}}
\frac{\Phi^c}{M_\P^c} 
(y y^\dagger)^a \frac{y^{(d+2-2a)}}{M_\P^d}, 
\ee
where negative $c$ can be taken to represent powers of $\overline{\Phi} $.
The condition $R=0$ becomes,
\be
d=2 (a+c-1),
\ee
so that only even-dimension operators may appear in $K$ as required.
In a fully viable model, one would also have to take account of
anomalies in the $R$ symmetry which can usually be cancelled if 
there are enough hidden sector singlets~\cite{herbi}. This will 
not be considered here. 

\section{Models with Duality Symmetry}
 
The second symmetry one can use to forbid terms linear in $N$ is
target space duality in a string effective action. 
Generally, these have flat directions, 
some of which correspond to moduli determining the size and shape of the
compactified space. Furthermore these moduli have discrete 
duality symmetries, which at certain points of enhanced symmetry become
continuous gauge symmetries~\cite{duality}. 
 
In Calabi-Yau models, abelian orbifolds and fermionic strings the
moduli include three K\"ahler class moduli ($T$-type) which are always
present, plus the possible deformations of the complex structure
($U$-type), all of which are gauge singlets. Additionally there
will generally be complex Wilson line fields~\cite{moduli,moduli2}.
When the latter acquire a vacuum expectation value they result in
the breaking of gauge symmetries. There has been continued interest
in string effective actions since they may induce the higgs
$\mu$-term~\cite{gm,moduli2,ant1,brignole}, be able to explain the
Yukawa structure~\cite{kpz,binetruy}, and be able to explain the
smallness of the cosmological constant in a {\em no-scale}
fashion~\cite{kpz,noscale}. Since the main objective here is 
simply to find a route to a viable low energy model with visible 
higgs singlets, these questions will only be partially addressed.

Typically the moduli and matter fields describe a space whose local 
structure is given by a 
direct product of $SU(n,m)/SU(n)\times SU(m)$ and $SO(n,m)/SO(n)\times
SO(m)$ factors~\cite{moduli,moduli2}. As an example consider
the K\"ahler potential derived in refs.\cite{moduli2}, which 
at the tree level is of the form 
\be
\label{stringkahler}
K=-\log (S+\overline{S}) -\log
[(T+\overline{T})(U+\overline{U})-\frac{1}{2}
(\Phi_1+\overline{\Phi}_2)(\Phi_2+\overline{\Phi}_1)]+\ldots
\ee
The $S$ superfield is the dilaton/axion chiral multiplet, and the
ellipsis stands for terms involving the matter fields. 
The fields $\Phi_1$ and $\Phi_2$ are two Wilson line moduli. As in
ref.\cite{gm,moduli2,ant1,brignole}, let us identify these fields with the neutral
components of the higgs doublets in order to provide a $\mu$-term. 
Problems such as how the dilaton acquires a VEV, or the eventual
mechanism which seeds supersymmetry breaking will not be addressed here. 

The moduli space is given locally by 
\be
{\cal{K}}_0 = \frac{SU(1,1)}{U(1)}\times \frac{SO(2,4)}{SO(2)\times
  SO(4)},
\ee
which ensures the vanishing of the scalar potential at least at the 
tree level, provided that the $S$, $T$ and $U$ fields all participate
in supersymmetry breaking (i.e. $G_S$, $G_T$, $G_U\neq 0$). 
In fact writing the K\"ahler function as
\be
G=K(z_i,z^i)+\ln \left| \hat{W}(z_i) \right|^2 ,
\ee
the scalar potential becomes 
\be
\hat{V}_s = - e^{G} \left(3- G_i G^{i\overline{j}} G_{\overline{j}} \right)
+ \frac{g^2}{2} {\rm Re} (G^i T^{Aj}_iz_j)(G^k T^{Al}_kz_l),
\ee
where $G_i=\partial G/\partial z^i$, and $G^{i{\overline{j}}}=
(G_{{\overline{j}}i})^{-1}$. The dilaton contribution separates, and
gives $G_S G^{S\overline{S}} G_{\overline{S}}=1$. To show that the
remaining contribution is $2$, it is simplest to define the vector
\be
A^\alpha= a (t,u,h,\overline{h})
\ee
where the components are defined as $\alpha=(1\ldots 4) \equiv
(T,U,\Phi_1,\Phi_2)$, and $u=U+\overline{U}$, $t=T+\overline{T}$,
$h=\Phi_1+\overline{\Phi}_2$. It is easy to show that
\be
G_\alpha A^\alpha =-2 a.
\ee
The vector $A^\alpha$ is designed so that $G_{\overline{\beta}\alpha} 
A^\alpha $ is proportional to $G_{\overline{\beta}}$; viz,
\be
G_{\overline{\beta}\alpha}A^\alpha = - a G_{\overline{\beta}}.
\ee 
Multiplying both sides by $G_\alpha G^{\alpha\overline{\beta}}$ gives
the desired result, i.e. that $G_\alpha G^{\alpha\overline{\beta}}
G_{\overline{\beta}}=2$. Thus, if the VEVs of the matter fields are
zero, the potential vanishes and is flat for all values of the moduli
$T$ and $U$, along the direction $\langle |\Phi_1|\rangle = \langle
|\Phi_2|\rangle=\rho_{\phi}$ (since this is the direction in which the
$D$-terms vanish).  The gravitino mass is therefore undetermined 
at tree level, being given by
\be
m^2 = \langle e^G \rangle = \frac{|\hat{W}|^2}{s(ut-2 \rho^2_{\phi})}.
\ee
In addition to the properties described above, there is an 
$O(2,4,Z)$ duality corresponding to automorphisms of the compactification 
lattice~\cite{duality,moduli2}. This constrains the possible form of the
superpotential. The $PSL(2,Z)_T$ subgroup implies 
invariance under the transformations~\cite{duality,moduli2},  
\ba
\label{dualtrans}
T &\rightarrow& \frac{aT-ib}{icT+d} \nonumber\\
U &\rightarrow& U-\frac{ic}{2}\frac{\Phi_1\Phi_2}{icT+d} \nonumber\\
z_i &\rightarrow& z_i (icT +d)^{n_i},
\ea
where $a,b,c,d~\epsilon~Z$, $ad-bc=1$, and 
where $z_i$ stands for general matter superfields with
weight $n_i$ under the modular transformation above. 
The $\Phi_1$ and $\Phi_2$ fields have modular weight $-1$. 
It is easy to verify the invariance of the K\"ahler function under
this transformation provided that 
\be
\label{wdual}
\hat{W}\rightarrow (ic T + d)^{-1} \hat{W}.
\ee
The superpotential should be defined to be consistent with this
requirement in addition to charge invariance, and this leads to a
constraint on the modular weights of the Yukawa couplings and matter
fields. (Anomalies occur here also, and must be cancelled in addition
to the gauge anomalies. Again this is considered to be beyond the
scope of the present paper.) 

One may now easily find examples where this symmetry is able 
by itself, to forbid dangerous operators. 
Consider the NMSSM superpotential of
eqn.(\ref{superpot2}). Identifying $\Phi_1$ and $\Phi_2$ with
the higgs superfields $H_1$ and $H_2$ (in order to generate a $\mu H_1
H_2 $ term in the low energy superpotential $W$) 
means that both of these fields have weight $-1$. 
Since the superpotential must transform as in eq.(\ref{wdual}), the 
other weights must obey the following;
\ba
3 n_N + n_k &=& -1 \nonumber\\
n_N   + n_\lambda &=& +1.
\ea
Since the Yukawa couplings are functions of the moduli, they too can
carry weight under the transformation in eqn.(\ref{dualtrans}).

One simple solution which forbids dangerous divergences 
is $n_N=-1$ and $n_k=n_\lambda=+2$. In this case it is obvious that 
(since the visible fields all have weight $-1$) even operators may be
avoided in $\hat{W}$. As for the K\"ahler potential, one 
expects the terms in $K_{\rm non-renorm}$ to be 
multiplied by powers of $(T+\overline{T})$. Thus terms in which the
holomorphic and anti-holomorphic weights are the same may be allowed.
Since all the weights are $-1$, this can obviously only be achieved for 
operators which have an even number of fields. 

There are clearly many ways in which one could devise similar models.
A perhaps more obvious example would be models in which the
superpotential transforms with weight $-3$. There all the physical
fields could be given weight $-1$, with the couplings having weight
$0$. It is then clear that only trilinear couplings can exist in
the superpotential, and only even-dimension terms can appear in the
K\"ahler potential.
  
\section{Conclusions} 

The problem of destablising divergences in models which extend the 
MSSM with a singlet field has been discussed. 
In this paper the case where there is no discrete or global symmetry 
at the weak scale has been examined, and the dangerously divergent 
tadpole diagrams have been identified. In particular it was shown 
that half of the possible operators (i.e. those with odd-dimension in the 
superpotential $\hat{W}$, or even-dimension in the K\"ahler potential) 
are perfectly harmless in the sense that they do not destroy the
gauged hierarchy. Thus an attractive possibility for
extending the higgs sector with a singlet is to generate the $\mu $ term from
couplings in the K\"ahler potential. Two examples were demonstrated in which 
all operators which are dangerous to the gauge hierarchy are
forbidden. In order to achieve this, they had to incorporate
either a gauged $R$-symmetry or a target space duality symmetry in
the full theory including gravity.   
These models clearly satisfy all constraints from fine-tuning,
primordial nucleosynthesis and cosmological domain walls.
Since they have no discrete or continuous 
global symmetries in the weak scale effective theories, one 
expects all possible couplings
(i.e. $\mu H_1 H_2 $, $\mu N^2 $, $\lambda N H_1 H_2 $ and $k N^3
$) to be present. The phenomenological 
implications of these more general cases, have been discussed recently 
in ref.\cite{moorehouse}. 

\vspace{1cm}
\noindent
{\bf \Large Acknowledgements:} I would like to thank H.~Dreiner,
J.-M.~Fr\`ere, M.~Hindmarsh, S.~King, D.~Lyth, G.~Ross, S.~Sarkar 
and P.~Van Driel for valuable discussions. This work was 
supported in part by the European Flavourdynamics Network
(ref.chrx-ct93-0132) and by INTAS project 94/2352.

\end{document}